%% file: Sinan_asplos21.tex
\pdfoutput=1
\documentclass[10pt]{jpaper}

\usepackage{etex}
\usepackage[sort,nocompress,adjust]{cite} 

\usepackage{amsmath,amssymb,amsfonts}
\usepackage{algorithmic}
\usepackage{graphicx}
\usepackage{listings}
\usepackage{xcolor}
\definecolor{olive}{rgb}{0.33, 0.42, 0.18}
\usepackage{makecell}
\usepackage{textcomp}
\usepackage{fancyhdr}
\usepackage{tabu}
\usepackage{enumitem}
\usepackage[normalem]{ulem}
\usepackage{inconsolata}
\usepackage[scaled]{beramono}
\usepackage{float}
\usepackage{caption}

\def\BibTeX{{\rm B\kern-.05em{\sc i\kern-.025em b}\kern-.08em
T\kern-.1667em\lower.7ex\hbox{E}\kern-.125emX}}

\usepackage{verbatim}   

\usepackage{wrapfig}
\usepackage{multirow}
\usepackage{balance}
\usepackage{cite}

\usepackage{pifont}

\newcommand{\smallcapital}{\fontsize{9pt}{10pt}\selectfont}

\usepackage{rotating}

\usepackage{arydshln}

\usepackage{placeins}

\usepackage{gensymb}

\usepackage{enumitem}
\usepackage[ruled,vlined]{algorithm2e}
\usepackage{upgreek,textgreek}
\usepackage[htt]{hyphenat}
\usepackage{appendix}

\hypersetup{colorlinks,
	linkcolor=blue,
	citecolor=blue,
	urlcolor=blue,
	filecolor=blue}

\usepackage{mathtools}

\usepackage{tikz}

\usepackage{listings}
\usepackage[utf8]{inputenc}
\lstdefinestyle{customcpp}{
	 aboveskip=0in,
	  belowskip=0in,
	   abovecaptionskip=0in,
	    belowcaptionskip=0in,
	     captionpos=b,
	      xleftmargin=\parindent,
	       language=C++,
	        morekeywords={forall},
		 showstringspaces=false,
		  basicstyle={\linespread{0.6}\fontseries{sb}\small\ttfamily},
		   keywordstyle=\bfseries,
		    commentstyle=\itshape\color{green!40!black},
	    }

\begin{document}

%
\title{Sage: Leveraging ML to Diagnose Unpredictable Performance in Cloud Microservices\vspace{-0.16in}}
%
\author{$^{1}$Yu Gan, $^{1}$Mingyu Liang, $^{2}$Sundar Dev, $^{2}$David Lo, and $^{1}$Christina Delimitrou\\ $^{1}$Cornell University, $^{2}$Google\\Contact author: yg397@cornell.edu}

\date{}

\maketitle
\pagestyle{plain}

\begin{abstract}
\input{Abstract.tex}
\end{abstract}

%
%
\input{Introduction.tex}

\input{Techniques.tex}

\input{Design.tex}

\input{Methodology.tex}

\input{Evaluation.tex}

\input{Conclusions.tex}

\section*{Acknowledgements}

We sincerely thank Daniel Sanchez and the anonymous reviewers for their feedback on earlier versions of this manuscript. This work was in part supported by NSF grants NeTS CSR-1704742, CCF-1846046, a Google Faculty Award, a Microsoft Research Faculty Fellowship, and a John and Norma Balen Sesquisentennial Faculty Fellowship. 

\balance
%

\bibliographystyle{IEEEtranS}
\bibliography{references}


\end{document}

%% file: Abstract.tex
Cloud applications are increasingly shifting from large monolithic services, 
to complex graphs of loosely-coupled microservices. Despite their advantages, microservices also introduce cascading QoS violations in cloud applications, which are difficult to diagnose and correct. 

We present \textit{Sage}, a ML-driven root cause analysis system for 
interactive cloud microservices. Sage leverages unsupervised learning models to circumvent the overhead of trace labeling, determines the root cause of unpredictable performance online, and applies corrective actions to restore performance. 
On experiments on both dedicated local clusters and large GCE clusters we show that Sage achieves 
high root cause detection accuracy and predictable performance.  

%% file: Introduction.tex
\section{Introduction}

Cloud computing has reached proliferation by offering \textit{resource flexibility}, \textit{cost efficiency}, and \textit{fast deployment}~\cite{BarrosoBook,Meisner11,Delimitrou14}. 
As the cloud scale and complexity increased, cloud services have undergone a major shift from large monolithic designs to complex graphs of single-concerned, loosely-coupled microservices. 
This shift is becoming increasingly pervasive, with large cloud providers, such as Amazon, Twitter, Netflix, and eBay having already adopted this application model~\cite{Cockroft15,Cockroft16,twitter_decomposing}. 
Microservices are appealing for several reasons, such as facilitating development, promoting elasticity, and enabling software heterogeneity. 

Despite their advantages, microservices also complicate resource management, as dependencies between tiers introduce backpressure, causing unpredictable performance to propagate through the system~\cite{Gan19,Delimitrou19}. 
Diagnosing such issues empirically is cumbersome and prone to errors, especially as 
typical microservices deployments include hundreds/thousands of tiers. Similarly, current 
cluster managers~\cite{Delimitrou13,Delimitrou13e,Delimitrou13d,Delimitrou14,Delimitrou15,Delimitrou16,Delimitrou17,Borg,Lin11,Meisner11,Lo14,Lo15,Mars13a,omega13,chen2019parties}. 
are not expressive enough to account for the impact of dependencies, putting more 
pressure on the need for automated root cause analysis systems. 

Over the past few years, there has been increased attention on trace-based methods to analyze~\cite{xtrace}, diagnose, and in some cases anticipate~\cite{Gan18b,Delimitrou19} performance issues in cloud services. 
While most of these systems target cloud applications, the only one focusing on microservices is Seer~\cite{Delimitrou19}, a DL-based system that anticipates cases of unpredictable performance by inferring the impact of outstanding requests on end-to-end performance. Despite its high accuracy, Seer leverages supervised learning to anticipate QoS violations, which require offline and online trace labeling. In a production system, this is non-trivial, as it involves injecting resource contention in live applications, hurting user experience. 





We present Sage, a root cause analysis system for interactive microservices that leverages unsupervised learning to identify the culprit of unpredictable performance in complex graphs of microservices. Sage does not rely on data labeling, hence it can be entirely transparent to both cloud users and application developers, scales well with the number of microservices and machines, and only relies on lightweight tracing that does not require application changes or kernel instrumentaion. We have evaluated Sage both on dedicated local clusters and large GCE settings and showed high root cause detection accuracy and improved performance predictability.


%% file: Techniques.tex
\section{ML for Performance Debugging}
\label{section:techniques}

\subsection{Overview}
Sage is a performance debugging and root cause analysis system for large-scale cloud applications. While the design centers around interactive microservices, where dependencies between tiers are more impactful, Sage is also applicable to traditional monolithic or SOA services. Sage relies on two broad techniques, each of which is described in detail below; first, it uses Causal Bayesian Networks ({\smallcapital CBN}) to model the RPC-level dependencies between microservices, and the latency propagation from the backend to frontend. Second, it uses a graphical variational auto-encoder ({\smallcapital GVAE}) to generate examples of counterfactuals~\cite{sep-causation-law}, and infer the hypothetical end-to-end latency had some occurring events not happened. Using these two techniques, Sage determines which set of microservices initiated an end-to-end QoS violation, and moves to adjust deployment and/or resource allocation to correct it. 


\subsection{Microservice Latency Propagation}
\label{sec:multiple_rpcs}

\begin{figure}[htb]
    \centering
    \includegraphics[width=0.42\textwidth]{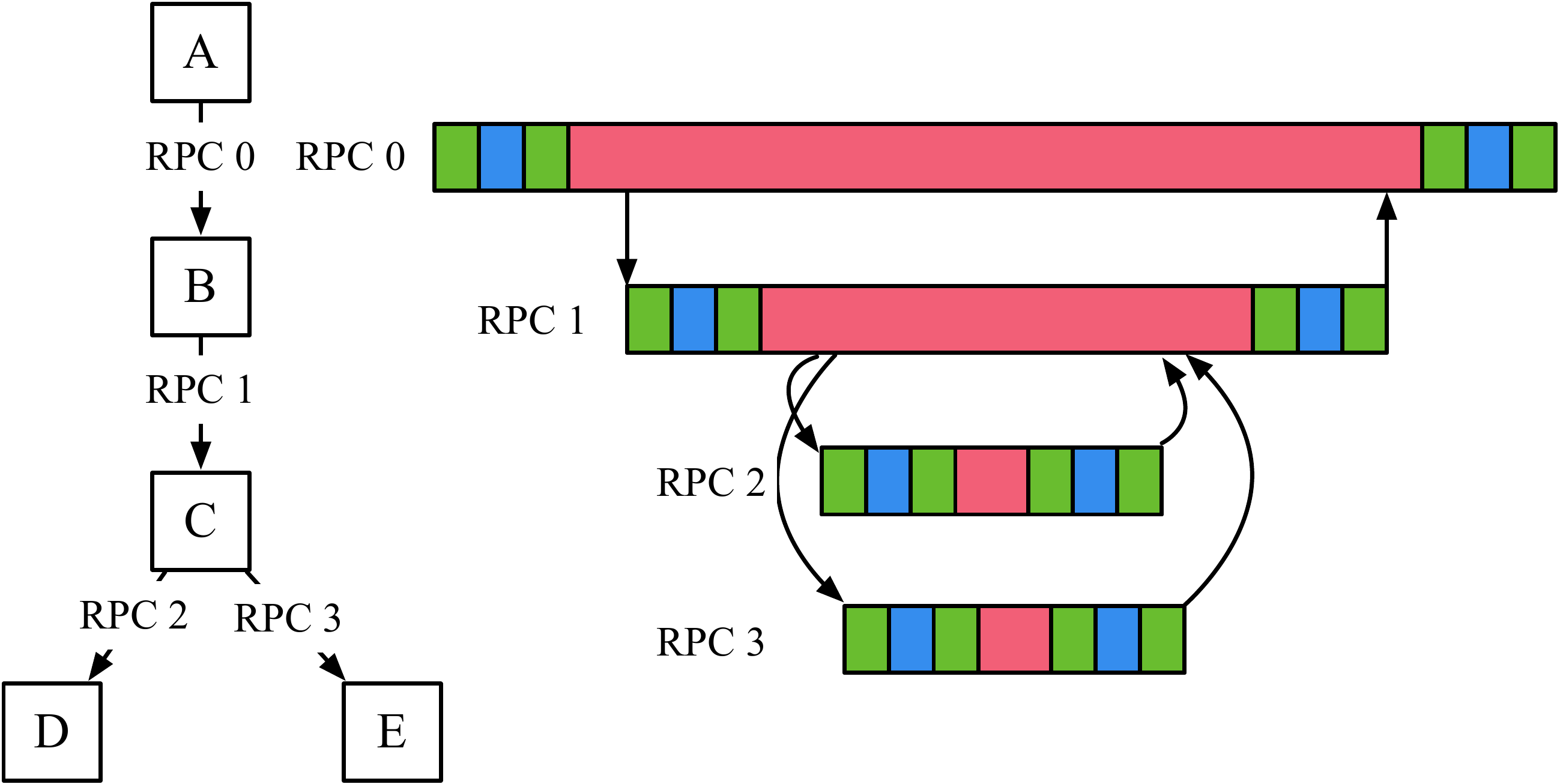}
    \caption{The dependency graph and traces of nested RPCs. }
    \label{fig:multiple_rpcs}
\end{figure}

Multiple RPCs between microservices form a tree of nested traces in a distributed monitoring system. Figure~\ref{fig:multiple_rpcs} shows
an example of the RPC dependency graph containing five services, four RPCs, and its corresponding latency traces.

The server-side latency of any non-leaf RPC is determined by the processing time of the RPC itself
and the waiting time (i.e., client-side latency) of its child RPCs. The latency of any RPC will propagate through the RPC dependency graph to the frontend and impact the end-to-end latency.
Since the latency of a child RPC cannot propagate to its parent without impacting its own latency,
the RPC latency propagation follows a \textit{local Markov property}, where each latency variable
is conditionally independent on its non-ancestor RPC latencies, given its child RPC latencies~\cite{koller2009probabilistic}.
For instance, the latency of \texttt{RPC0} is conditionally independent on the latency of \texttt{RPC2}
and \texttt{RPC3}, given the latency of \texttt{RPC1}. 

\subsection{Modeling Microservice Dependency Graphs}

Causal Bayesian Networks (CBN) are a common tool to model causal relationships~\cite{neapolitan2004learning,pearl2009causal}.
A CBN is a directed acyclic graph (DAG), where the nodes are different random variables
and the edges indicate their conditional dependencies. More specifically, each edge represents
a causal relationship, directed from the cause to the effect. We define three types of nodes in the Bayesian network:
\begin{itemize}
    \item \textbf{Metric nodes ($X$)}: The metric nodes contain primarily resource-related metrics of all services and network channels, such as CPU and memory utilization and network bandwidth, 
          collected via the in-place monitoring tools~\cite{google-cloud-monitoring, aws-cloudwatch, azure-monitoring}.
    \item \textbf{Latency nodes ($Y$)}: The latency nodes include client-side latency ($Y^c$), server-side latency ($Y^s$),
          and request/response network delay ($Y^{req}$ and $Y^{resp}$) of all RPCs. 
    \item \textbf{Latent variables ($Z$)}: The latent variables contain the unobserved and immeasurable factors that are responsible for latency stochasticity. They are critical to generate the counterfactual latencies Sage relies on to diagnose root causes of QoS violations. 
\end{itemize}

We can construct the CBN among the three-node
classes of all RPCs based on the inherent causal relationships and latency propagation observations
obtained via a distributed tracing system, such as Dapper or Zipkin.

\begin{wrapfigure}[12]{r}{0.29\textwidth}
    \centering
\vspace{-0.2in}
    \includegraphics[scale=0.33, viewport=60 36 400 430]{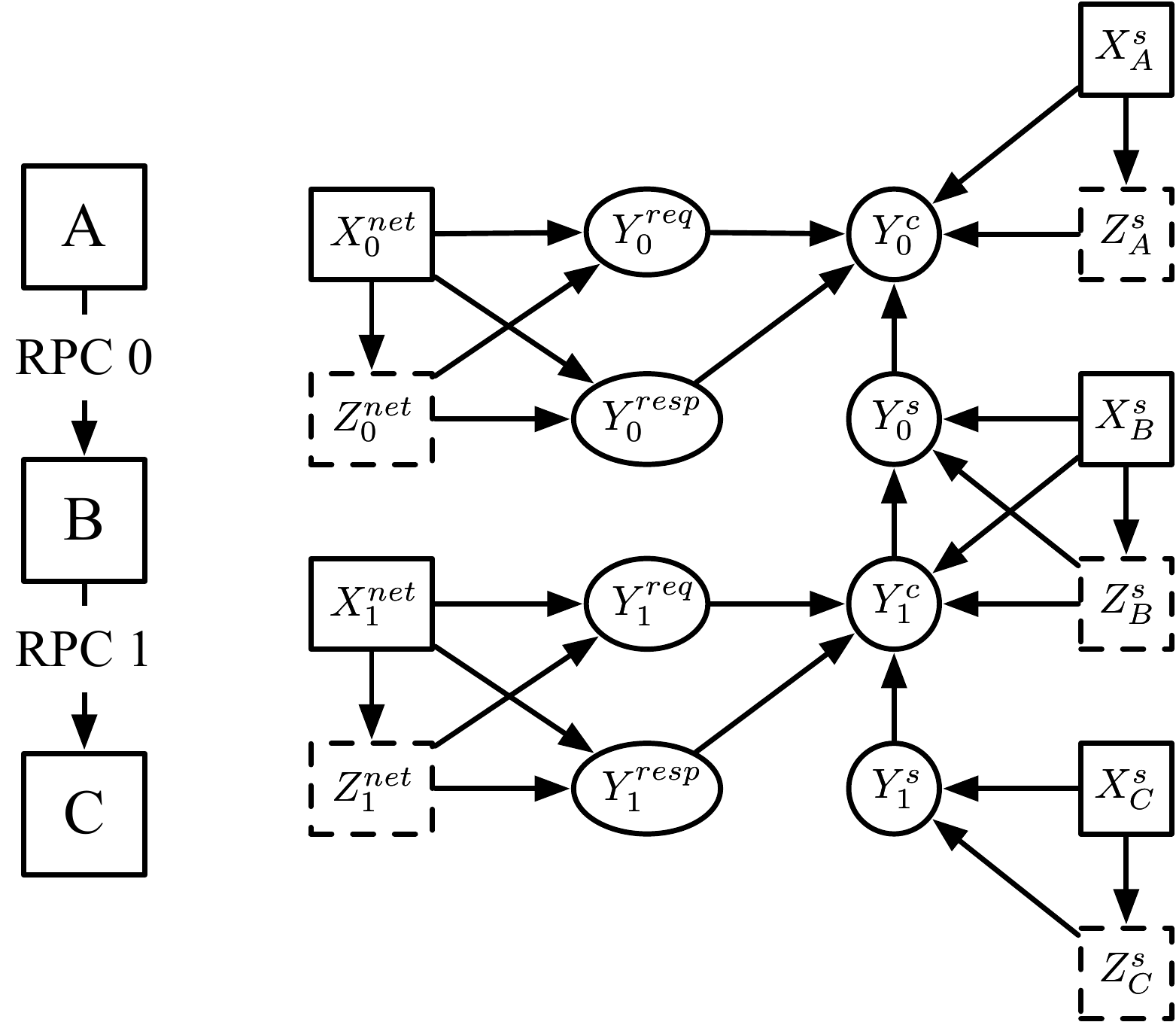}
    \caption{Latency propagation CBN. }    \label{fig:cbn_two_rpcs}
\end{wrapfigure}

Figure~\ref{fig:cbn_two_rpcs} shows an example of the CBN constructed for a three-microser-vice chain based on the RPC dependency graph.
The nodes with solid lines ($X$ and $Y$) are observed, while the nodes with dashed lines ($Z$) are latent variables that need to be inferred. The arrows in the RPC dependency graph and CBN have opposite directions because the latency of one RPC is determined by the latency of its child RPCs.

\subsection{Counterfactual Queries}

In a typical cloud environment, site reliability engineers (SREs) can verify if a suspected root cause is correct by reverting a microservice's configuration to a state known to be safe, while keeping the remaining microservices unchanged. If the problem is resolved, the suspected culprit is causally related to the QoS violation. Sage uses a similar process, where ``suspected root causes'' are generated using counterfactuals, which determine the causal effect by asking what the outcome would be if the state of a microservice had been different~\cite{morgan2015counterfactuals, menzies2008counterfactual, hofler2005causal}.

To avoid impacting the performance of live services during this process, Sage leverages historical tracing data to generate realistic counterfactuals~\cite{pearl2009causal, morgan2015counterfactuals}, taking into account that the exact situation may not have occurred in the past. If the probability that the end-to-end tail latency meets QoS after intervention is greater than a threshold, then those services are the root cause of the performance issue.

Conditional deep generative models, such as the conditional variational autoencoders (CVAE)~\cite{NIPS2015_5775}
and conditional generative adversarial nets, are common tools to generate new data of a class from an original distribution. Generally, they compress a high-dimensional target ($Y$) and tag ($X$) into low-dimensional latent space variables ($Z$), and use the latent space variables and tags to generate new target data. Recent studies have showed that these techniques can also be used to generate counterfactuals for causal inference~\cite{louizos2017causal,yoon2018lifelong}. 

To generate counterfactuals, we build a network of CVAEs according to the structure of the CBN. Although using one CVAE for the entire microservice graph would be simple, it has several drawbacks. First, it lacks the CBN's structural information which is useful in terms of avoiding ineffectual counterfactuals based on spurious correlations. Second, it prohibits partial retraining and transfer learning, which is essential given the frequent update cadence of microservices. Finally, the black-box model is less explainable since it cannot reveal any information on how the latency of a problematic service propagates to the frontend.
Therefore, we construct one lightweight CVAE per microservice, and connect the different CVAEs according to the structure of the CBN to form the graphical variational autoencoder (GVAE).

The encoders and prior networks take the observed metrics as inputs, and are trained in parallel. The decoders require the outputs of the parent decoders in the CBN as inputs, and are trained serially. The maximum depth of the CBN determines the maximum number of serially-cascaded decoders.


%% file: Design.tex
\section{Sage Design}
\label{sec:design}


\begin{figure}
  \centering
  \includegraphics[width=0.46\textwidth]{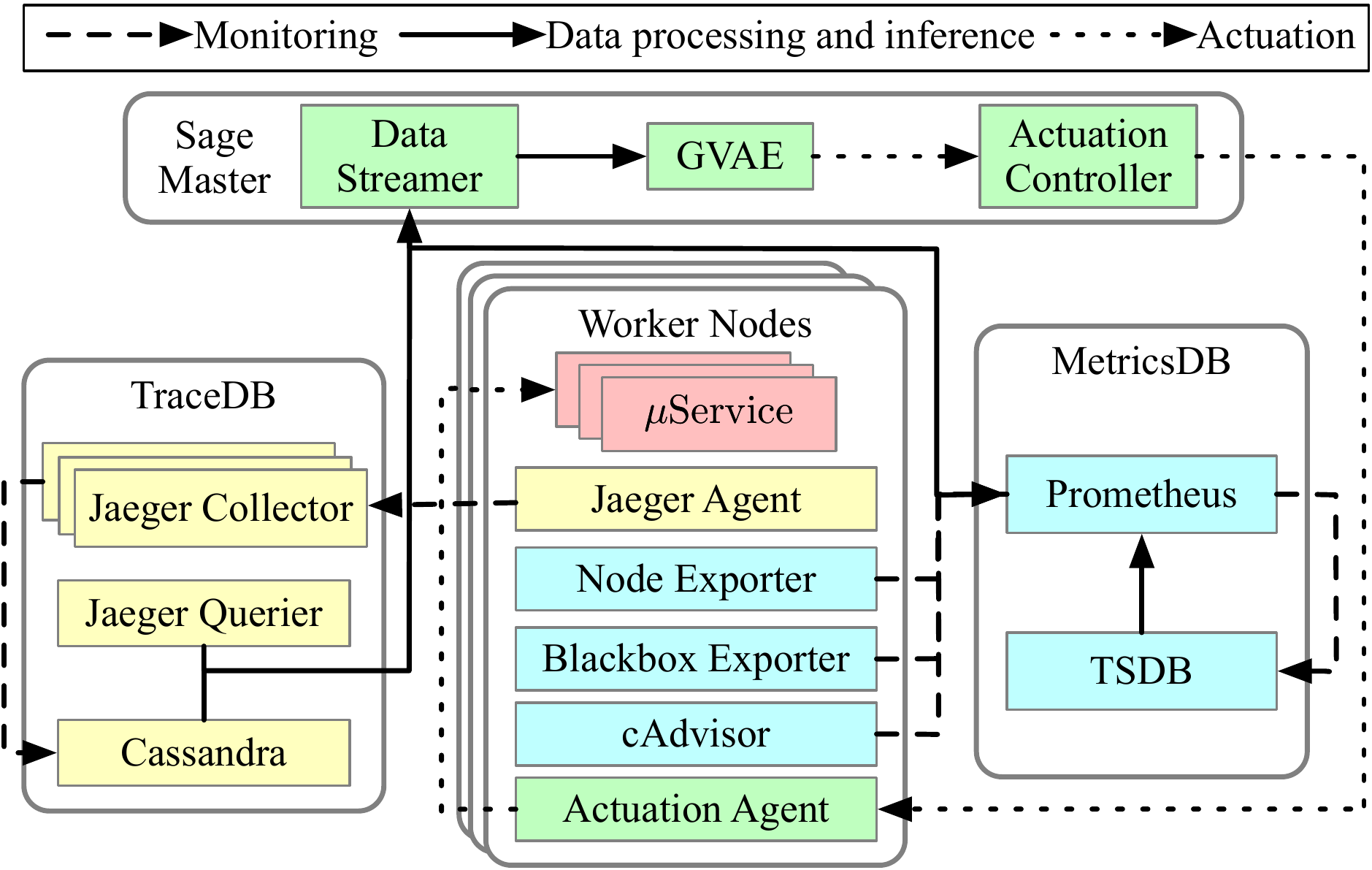}
  \caption{Overview of Sage's system design. }
  \label{fig:sage_overview}
\end{figure}

Sage is a root cause analysis system for complex graphs of interactive microservices. Sage relies on RPC-level tracing to compose the Causal Bayesian network of the microservice topology, and per-node tracing to track the per-tier latency distribution.
Below we discuss the training and inference pipeline (Sec.~\ref{sec:root_cause}), Sage's actuation system once a root cause has been identified (Sec.~\ref{sec:intervention}), and the way Sage handles changes in application design (Sec.~\ref{sec:retraining}). 

Fig.~\ref{fig:sage_overview} shows an overview of Sage. The system uses Jaeger~\cite{jaeger}, a distributed RPC-level 
tracing system to collect end-to-end request execution traces. Jaeger uses sampling to reduce tracing oveheads. Sage also uses the Prometheus Node Exporter~\cite{nodeexporter}, Prometheus Blackbox Exporter \cite{blackboxexporter}, and cAdvisor \cite{cadvisor} to collect machine- and container-level hardware/OS metrics, and network latencies. Each metric's timeseries is stored in the Prometheus TSDB~\cite{prometheus}. At runtime, Sage queries Jaeger and Prometheus periodically to obtain real-time latency and usage metrics. The GVAE then uses this data to infer the root cause of any QoS violation(s). Once a root cause is diagnosed, Sage's actuator takes action to restore performance by adjusting the offending microservice's resources. 

Sage uses a centralized master for trace processing and root cause analysis, and per-node agents for trace collection and container deployment. It also maintains two hot stand-by copies of the Sage master for fault tolerance. 


\subsection{Root Cause Analysis}
\label{sec:root_cause}

Sage first uses the Data Streamer to fetch 
and pre-process the latency and usage statistics. Sage initializes and trains the GVAE model offline with all initially available data. It then periodically retrains the model, even when there are no changes in application design, to account for changes in user behavior~\cite{parisi2019continual,hoi2018online,yoon2018lifelong}. Online learning models are prone to \textit{catastrophic forgetting}, where the model forgets previously-learned knowledge upon learning new information~\cite{parisi2019continual,kirkpatrick2017overcoming}. To avoid this, we interleave the current and previous data in the training batches. In addition, to avoid \textit{class imbalance}, i.e., cases where the datapoints that meet QoS are significantly more than those which violate it, the model oversamples the minority classes to create a more balanced training dataset. 

At runtime, Sage uses the latest version of the GVAE model to infer the root cause of QoS violations. Sage first calculates the medians of usage and performance metrics where QoS is met, and labels them as \textit{normal values}. If at any point QoS is not met, the GVAE will generate counterfactuals by replacing a microservice's metrics with their respective \textit{normal values}. The service whose counterfactual would resolve the QoS violation is identified as the culprit behind it. 



Sage implements a two-level approach to locate the root cause of a QoS violation. 
It first uses service-level counterfactuals to locate the culprit microservices, and then uses metric-level counterfactuals of the offending service to identify the underlying reason that caused it to become the culprit. 

\subsection{Actuation}
\label{sec:actuation}

Once Sage determines the root cause of a QoS violation it takes action. Depending on which resource is identified by the GVAE as the one instigating the QoS violation, Sage will dynamically adjust the CPU frequency, scale up or scale out the problematic microservices, limit the rate of the collocated interference jobs, partition the last level cache (LLC) with Intel Cache Allocation Technology (CAT), and partition the network bandwidth with Linux traffic control queuing disciplines. Sage first tries to resolve the performance issue by only adjusting resources on the offending node, and only when that is insufficient it scales out the problematic microservice on new nodes and/or migrate it. 

\subsection{Handling Microservice Updates}
\label{sec:retraining}

Training the complete model from scratch for large clusters takes tens of minutes to hours, so it is impractical to happen for every change in application design/deployment. Sage instead implements \textit{selective partial retraining} and \textit{incremental retraining} with a dynamically reshapable GVAE similar to~\cite{yoon2018lifelong}, thanks to VAE's ability to be decomposed using the CBN. 
On the one hand, with selective partial retraining, we only retrain the neurons corresponding to the updated nodes 
and their descendents in the CBN, because the causal relationships guarantee that all other nodes are not affected by the change. On the other hand, with incremental retraining, we initialize the parameters of the network to those of the previous model, while adding/removing/reshaping the corresponding networks if microservices are added/dropped/updated. 
The combination of these two \textit{transfer learning} approaches reduces retraining time by more than $10\times$, especially when there is large fanout in the RPC dependency graph. 

%% file: Methodology.tex
\section{Methodology}
\label{sec:methodology}

\subsection{Cloud Services}

\textbf{Generic Thrift microservices: } \textit{Apache Thrift}~\cite{thrift} is a popular RPC framework. We implement a code generator for composable Thrift microservices, and evaluate two topologies; a \textit{Chain} and a \textit{Fanout}. In Chain, each microservice receives a request from its parent, sends it to its downstream service, and responds to its parent once it gets the results from its child. In Fanout, the root service broadcast each request to all leaf services and returns to the client only after all children have responded. 
Most real microservice architectures are combinations of these two topologies~\cite{GrandSLAm,Gan19,usuite}.

\noindent{\textbf{Social Network: }} One of the end-to-end microservice in the DeathStarBench 
suite\cite{Gan19} that implements a broadcast-style social network with uni-directional follow relationships. 

\vspace{-0.08in}
\subsection{Systems}
\noindent{\textbf{Local Cluster: }} We use a dedicated local cluster with five 2-socket 40-core servers with 128GB RAM each, and two 2-socket 88-core servers with 188GB RAM each. Each server is connected to a 40Gbps ToR switch over 10Gbe NICs.

\noindent{\textbf{Google Compute Engine: }} We also deploy the Social Network service to Google Compute Engine (GCE)
with 84 nodes in \texttt{us-central1-a} to study Sage's scalability. All nodes are dedicated, so there is no interference from external jobs. 

\vspace{-0.08in}
\subsection{Training Dataset for Validation}

We use wrk2~\cite{wrk2} as the open-loop HTTP workload generator for all three applications. To verify the ground truth during validation, 
we use \texttt{stress-ng}~\cite{stress-ng} and \texttt{tc-netem}~\cite{tc-netem} to inject CPU-, memory-, disk-, and network-intensive microbenchmarks to different, randomly-chosen subsets of microservices. 

%% file: Evaluation.tex
\section{Evaluation}
\label{sec:evaluation}

\begin{figure}
  \centering
  \includegraphics[width=0.5\textwidth]{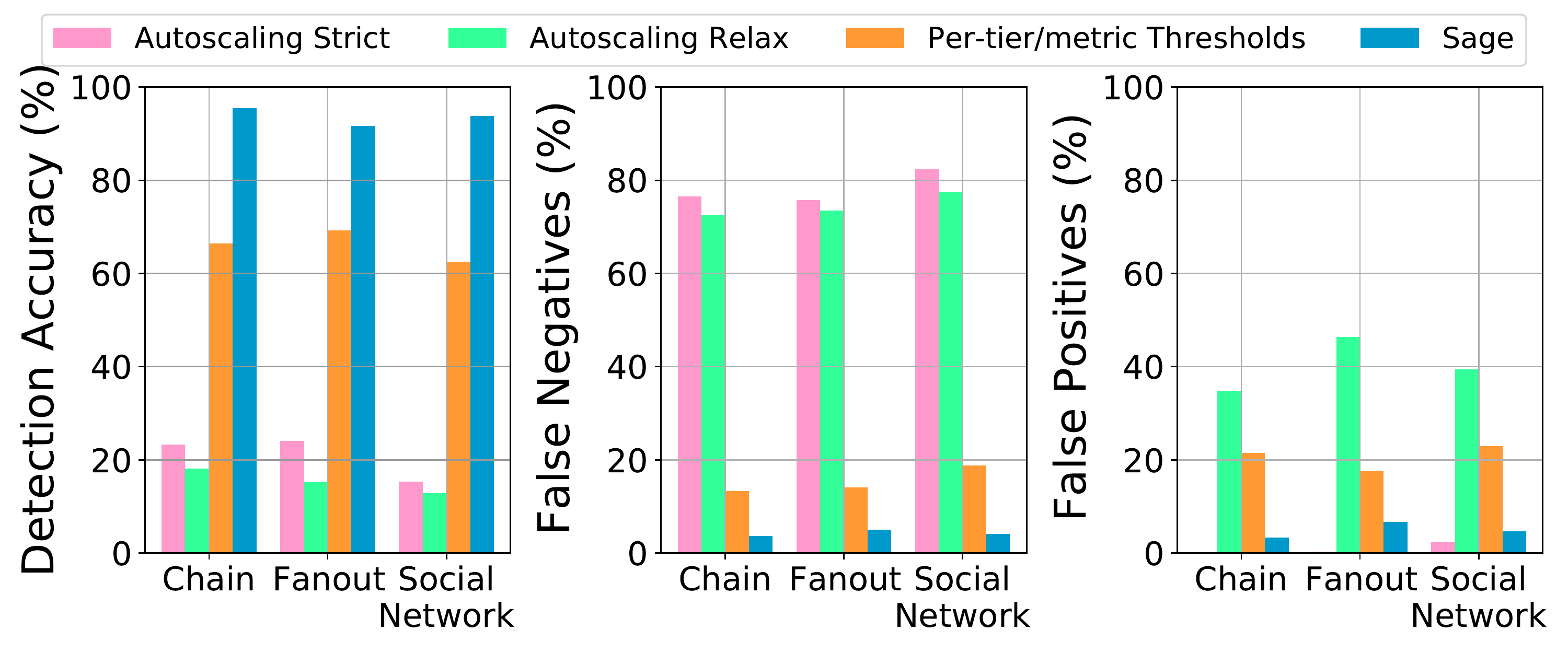}
  \vspace{-0.24in}
  \caption{Detection accuracy and false negatives/positives. }
  \label{fig:accuracy}
\end{figure}

\subsection{Sage Validation}
\label{sec:validation}

Fig.~\ref{fig:accuracy} shows the accuracy of Sage in the local cluster across services, compared to two autoscaling techniques, and an oracular scheme that sets thresholds for each tier and metric offline, beyond which point resources are upscaled. \textit{Autoscale Strict} increases resource allocation when the utilization of a microservice exceeds 50\% and \textit{Autoscale Relaxed} when it exceeds 70\% (on par with AWS's autoscaling policy). Sage significantly outperforms the other methods, even the offline oracular one, by learning the impact of dependencies between neighboring microservices. Similarly, Sage's false negative and false positive rates are marginal, which avoids QoS violations and resource inefficiencies respectively. 

\begin{figure}[htb]
  \centering
  \includegraphics[width=0.48\textwidth]{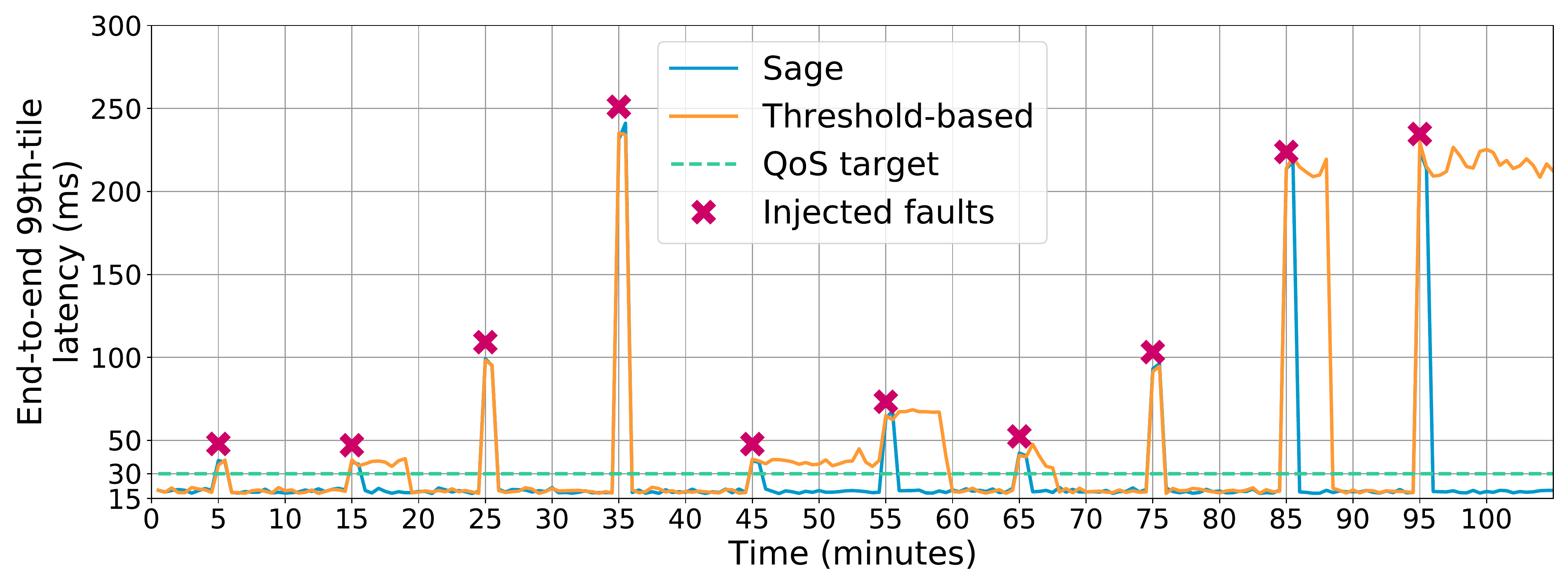}
  \vspace{-0.24in}
  \caption{End-to-end latency for Social Network in the presence of QoS violations. }
  \label{fig:intervention}
\end{figure}
\vspace{-0.08in}

\subsection{Actuation}
\label{sec:intervention}

Fig.~\ref{fig:intervention} shows the impact of root cause analysis on the end-to-end tail latency of the Social Network, with Sage and the offline oracular technique. To create unpredictable performance whose source is known, we inject resource intensive kernels in a randomly-selected subset of microservices. While there are cases that the threshold-based scheme identifies correctly, more often performance takes a long time to recover. In comparison, Sage immediately identifies the correct root cause, and applies corrective action to restore performance. 

We have also introduced changes to several microservices and have validated that the transfer learning in Sage reduces training time by at least an order of magnitude compared to retraining from scratch, without impacting accuracy.

\subsection{Scalability}
\label{sec:scalability}

 We now evaluate Sage's accuracy when deploying the Social Network on 188 instances on GCE using Docker Swarm. We replicate the stateless microservices and shard the caching systems and databases across instances during periods of higher load. Each service has 1-10 replicas, depending on the maximum single-process throughput. Accuracy on GCE is within 1\% of the detection accuracy on the local cluster, while the difference in false positives and false negatives is also marginal. 
 We also evaluate the difference in training and inference time between the local cluster and GCE. We use two Intel Xeon 6152 processors with 44 cores in total for training and inference. Training from scratch takes 124 minutes
for the local cluster and 148 minutes for GCE. Inference takes 49ms on the local cluster and 62ms on GCE. Although we deploy 6.7x more containers on GCE than on the local cluster,
the training and inference time only increase by 19.4\% and 26.5\% respectively.



%% file: Conclusions.tex
\section{Conclusions}
\label{sec:conclusions}

We have presented Sage, an ML-driven root cause analysis system for interactive, cloud microservices. Sage leverages entirely unsupervised models to detect the sources of unpredictable performance, removing the need for empirical diagnosis or expensive data labeling. Sage adapts to frequent design changes, and takes action to restore QoS. We shows that Sage achieves high root cause detection accuracy and improved performance. Given the increasing complexity of cloud services, data-driven systems like Sage can improve performance predictability without sacrificing efficiency.